# Ultrathin $Ga_2O_3$ Tunneling Contact for 2D Transition-metal Dichalcogenides Transistor

Yun Li [1,2], Tinghe Yun [1], Bohan Wei [1,3], Haoran Mu [1], Luojun Du [2], Nan Cui [1#], Guangyu Zhang [1,2#], Shenghuang Lin [1#]

1. Songshan Lake Materials Laboratory, Dongguan 523808, P. R. China.

2. Institute of Physics, Chinese Academy of Science, Beijing, 100190, P. R. China.

3. MOE Key Laboratory of Laser Life Science & Guangdong Provincial Key Laboratory of Laser Life Science, College of Biophotonics, South China Normal University, Guangzhou, 510631, China

These authors contributed equally: Yun Li, Tinghe Yun

[#]Corresponding author email:

    Cuinan@sslab.org.cn

    zhangguangyu@sslab.org.cn

    lingshenghuang@sslab.org.cn

# Abstract

The development of two-dimensional (2D) transition metal dichalcogenides (TMDs) based transistors has been constrained by high contact resistance and inadequate current delivery, primarily stemming from metal-induced gap states and Fermi level pinning. Research into addressing these challenges is essential for the advancing 2D transistors from laboratory experiments to industrial-grade production. In this work, we present amorphous $Ga_2O_3$ as a novel tunneling contact layer for multilayer $WS_2$-based field-effect transistors (FETs) to enhance electrical performance. The addition of this innovative tunneling layer avoid Schottky barrier forming while finally change into a tunneling barrier with the barrier height to just 3.7 meV, near-ideal ohmic contacts. This approach effectively reduces contact resistance to only 2.38 kΩ·μm and specific contact resistivity as low as 3 x $10^{-5}$ Ω·$cm^2$. A record-high electron mobility of 296 $cm^2$ $V^{-1}$ $s^{-1}$ and ON-OFF ratio over $10^6$ are realized for $WS_2$ transistor at room temperature. Compared to other tunneling materials, ultrathin $Ga_2O_3$ layer offers scalability, cost-efficient production and broad substrate compatibility, making it well-suited for seamless integration with industrial wafer-scale electronics. A robust device performance remains highly consistent in a large-scale transistor array fabricated on 1.5 × 1.5 cm² chips, with the average mobility closing to 200 $cm^2$ $V^{-1}$ $s^{-1}$. These findings establish a new benchmark for contact performance in 2D transistors and prove the potential of tunneling contact engineering in advancing high-performance, scalable electronics with promising applications in quantum computing and communication.

# Introduction

Two-dimensional (2D) transition metal dichalcogenides (TMDs), have emerged as leading candidates for advanced electronics and optoelectronics, owing to their high carrier mobility, tunable band gaps, and exceptional van der Waals (vdW) integrated characteristic [1,2]. However, achieving high-quality electrical contacts remains one of the key challenges in realizing the full potential of 2D field-effect transistors (FETs) for practical applications [3–6]. Conventional metal deposition techniques often result in Fermi level pinning effect at the TMD surface, which exacerbates contact resistance and prevents effective tuning of the Schottky barrier height (SBH) with different metals [7,8]. Large Schottky barrier will impede charge injection and transport, ultimately degrading device performance [9]. Various strategies focusing on the electrodes design of metal-semiconductor (MS) structure to mitigate Schottky barrier and reduce contact resistance, for examples, applying semimetal electrodes or vdW transfer electrodes [10–14]. Different from MS structure, inserting an ultrathin insulator layer to form a metal–insulator–semiconductor (MIS) structure has proven effective in avoiding Schottky barriers via quantum tunneling meanwhile the contact mode will be dominated by the tunneling resistance [15]. Atomic-layer deposition (ALD) of ultrathin oxide insulator such as $Ta_2O_5$ has been firstly explored for this purpose, but maintaining film continuity and compactness on TMD surfaces is normally difficult can be challenging due to the dangling bond-free 2D materials surface [16,17]. Alternative approaches have utilized 2D crystalline materials like hexagonal boron nitride (hBN), tungsten disulfide ($WS_2$), and multilayer InSe as tunneling layers for TMD-based FETs [18–20]. The vdW stacking fabrication of these materials avoided impact from material deposition, led to significant improvements of electrical performance [21]. Although current tunneling materials have shown potential in improving transistor performance, their ability to enhance key parameters, such as mobility, remains limited and not yet sufficient for meeting the demands of industrial applications. Additionally, the scalable fabrication of these materials often introduces constraints on substrate compatibility and processing temperatures, posing challenges for seamless integration with 2D transistors and industrial-grade production of heigh-performance devices[5,22,23].

To address these limitations, we explore the use of ultrathin layer $Ga_2O_3$, derived from the surface of liquid gallium, as a novel tunneling contact layer. Liquid gallium cannot be completely oxidized in ambient hence represented as $Ga_2O_3$. This dense, amorphous layer, with a scale of more than centimeter and thickness of 2-4 nm, possess mechanical flexibility, environmental stability and, most importantly, scalability [24].

Additionally, the compact Ga$_2$O$_3$ film acts as a passivation layer, protecting the functional layer from Fermi level pinning and providing robust encapsulation to isolate environmental contaminants such as water and oxygen molecules [25,26]. It can also protect transistor from external charge injection and electromagnetic interference while reduce the damage of electrostatic discharge [27,28]. Multilayer WS$_2$ offers significant advantages for transistor applications, including high electronic mobility, tunable bandgap and robust thermal stability. Previous studies have also demonstrated the encapsulation effect of Ga$_2$O$_3$ on WS$_2$ that enhances optical performance of WS$_2$ [29]. Building on these findings, we focused on multilayer WS$_2$ transistors as a typical case to investigate the performance improvements enabled by Ga$_2$O$_3$ tunneling contacts.

We successfully demonstrate the incorporation of ultrathin Ga$_2$O$_3$ as tunneling contact layers in multilayer WS$_2$ (<10 layers) transistor and observe a dramatic reduction in barrier height from 258 meV to just 3.7 meV. Notably, the transistor approaches the ideal of ohmic contact since the contact resistance of the transistor is largely influenced by the tunneling resistance. The final specific contact resistivity ($\rho_c$) falls within the range of 3-25 x 10$^{-5}$ Ω·cm$^2$. Highest carrier mobility up to 296 cm$^2$ V$^{-1}$ s$^{-1}$ at room temperature is achieved with tunneling contact, which is a record-high value for multilayer WS$_2$-based transistors, to the best of our knowledge [21]. Our results also show consistent performance of a transistor array across a 1.5 × 1.5 cm$^2$ chip, attributed to the high quality and uniformity of large-area printed ultrathin Ga$_2$O$_3$ layer. These findings demonstrate the exceptional potential of ultrathin amorphous Ga$_2$O$_3$ as an effective tunneling contact layer for 2D semiconductor devices. By addressing both performance and scalability challenges, this strategy enhances the practical application and commercial viability of 2D semiconductor technologies.

## Results

**Large-scale contact layer preparation and characterization**

Ultrathin Ga$_2$O$_3$ layer was synthesized by printing of native oxide onto a variety of substrates [30]. The printing installation and oxide printing process of Ga$_2$O$_3$ tunneling layer was illustrated in **Fig. 1a**. The polypropylene carbonate (PPC) spin-coated on polydimethylsiloxane (PDMS) serves as the desired substrate. The liquid metal forms an oxide-coated meniscus that bridges the gap between the printer head and the substrate. As the substrate moves, an ultrathin $Ga_2O_3$ layer is deposited with minimal residual metal. The film was then transferred on top of $WS_2$ nanoflakes with the association of PPC as detailed in Supplementary **Fig. S1** and the supporting notes. **Fig. 1b** demonstrates a 3D diagram of tunneling contact transistor assembly structure, while gold (Au) electrodes are insulated from $WS_2$ nanoflake by the ultrathin $Ga_2O_3$ layer. In the tunneling contact device, current will firstly be tunneling beyond the tunneling layer and then transfer in channel material when a bias voltage is applied on source and drain electrodes.

To facilitate a direct comparison within a single device, Au electrodes were deposited on both the $Ga_2O_3$ and $WS_2$ surfaces, with a channel length ($L_{CH}$) of 4.5 μm and width ($L_{CW}$) of 20 μm, as depicted in the optical image in **Fig. 1c** (top). The ultrathin $Ga_2O_3$ layer formation is characterized as a self-limiting oxidation process, producing a dense film that protects the transistor channel from environmental factors and prevents further oxidation [26]. Supplementary **Fig. S2a** displayed photographs of ultrathin $Ga_2O_3$ layer printed on PDMS and PPC surface and then transferred on $SiO_2$/Si substrates, highlighting the versatility and adaptability of the process. These photos showcased the uniformity and continuity of $Ga_2O_3$ layers with no-selectivity of substrate, essential for achieving large-area transistor arrays on chips.

Despite variations in film thickness due to different oxidation time in ambient as well as synthesis methods, no correlation with device performance was observed. Specifically, the $Ga_2O_3$ film measured 2.6 nm in thickness for tunneling layer (**Fig. 1c**, bottom). The $WS_2$ nanoflakes is less than 10 layers (Supplementary **Fig. S2b**). The printed film has a metallic character that can oxidize over time and finally the gallium oxide tunneling layers with stoichiometric ratios approaching 2:3 can be formed, which was verified by X-ray photoelectron spectroscopy (XPS) test (Supplementary **Fig. S2c**). The XPS results confirm the successful formation of $Ga_2O_3$ with gallium predominantly in the $Ga^{3+}$ oxidation state. The Ga 3d peak (~20.0–21.0 eV) and Ga 2p peak (1117–

1118 eV) both exhibit symmetric shapes, indicating the absence of metallic Ga or other lower oxidation states. In the O 1s spectrum, the primary peak (530.1–531 eV) corresponds to lattice oxygen ($O^{2-}$) in the $Ga_2O_3$ structure. These results confirm the high purity and stable composition of the $Ga_2O_3$ layer, with only minimal surface contamination.

Raman spectra of the $WS_2$ nanoflake, shown in **Fig. 1d**, identified primary peaks at 2LA(M), E″(M), $E_{2g}^1$(Γ), and $A_{1g}$ (Γ) represent for the longitudinal acoustic mode and in-plane vibrational mode at the M point, in-plane vibrational mode and out-of-plane vibrational mode at the Gamma point from $WS_2$, respectively. The consistency in peak positions and full-width at half-maximum (FWHM) before and after $Ga_2O_3$ encapsulation indicated negligible strain or doping introduced during the process. Photoluminescence (PL) spectra (**Fig. 1e**) revealed a slight increase in exciton energy for $Ga_2O_3/WS_2$, with a 0.012 eV shift, attributed to dielectric screening from $Ga_2O_3$ film. These results confirmed the effective encapsulation of $Ga_2O_3$ layer, demonstrating its ability to maintain the stability and performance of $WS_2$-based transistors potential for reliable protection in electronic applications.

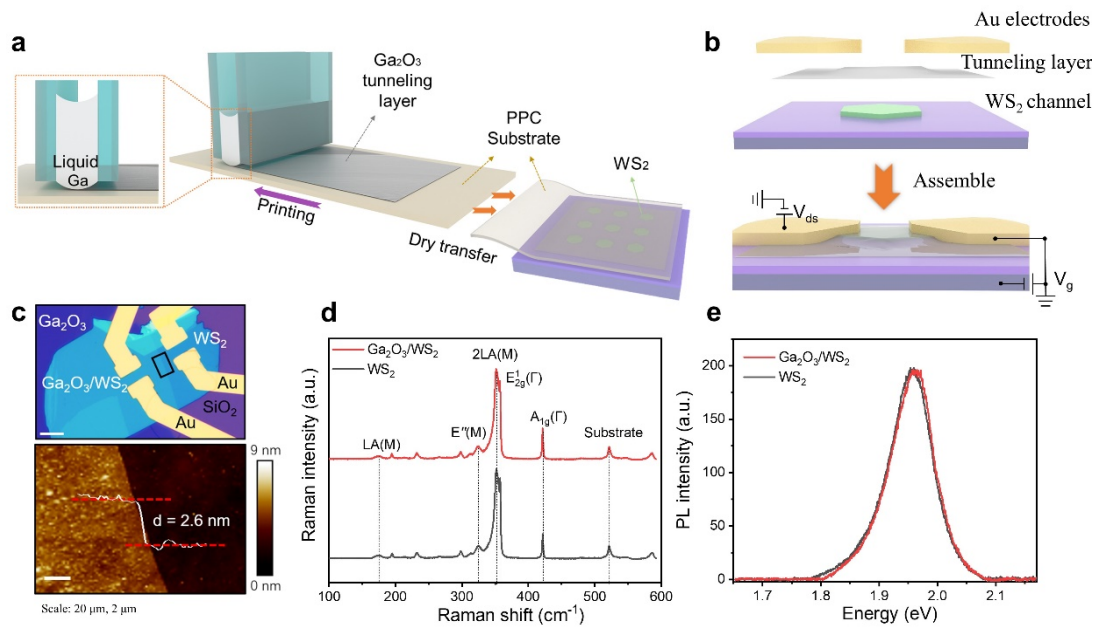

**Fig. 1| Device Structure and Characterization** (**a**) Schematic image of printing setup illustrating the continuous ultrathin $Ga_2O_3$ tunneling layer on to PPC-coated PDMS

substrate. (**b**) 3D diagram of tunneling contact transistor assembly structure. Ultrathin $Ga_2O_3$ tunneling layer separates the gold electrodes from the $WS_2$ channel material. $SiO_2$ (300 nm thickness) work as the dielectric layer. (**c**) Optical micrographs of the $Ga_2O_3/WS_2$ tunneling contact transistor (top) and atomic force micrographs (AFM) of $Ga_2O_3$ (bottom), showing a thickness of $Ga_2O_3$ layer is 2.6 nm. Scale bar: 20 μm (top) and 2 μm (bottom). (**d**) Raman spectra of both the $WS_2$ and $Ga_2O_3/WS_2$ regions, exhibiting no obvious wavenumber shift. (**e**) PL spectra of the $WS_2$ and $Ga_2O_3/WS_2$ regions with 0.01 eV peak shift.

**Enhanced electrical performance of $WS_2$ FETs with $Ga_2O_3$ tunneling layer**

To evaluate the potential of the $Ga_2O_3$ tunneling layer in enhancing the performance of $WS_2$-based FETs, electrical tests were conducted under various conditions. **Fig. 2a** presents the typical room-temperature transfer curves for a $Ga_2O_3/WS_2$ transistor tested in vacuum as the drain–source current ($I_{ds}$) changed with drain-source bias ($V_{ds}$) varying from 0.01 V to 1 V. An ultra-high electronic mobility ($\mu_e$) could achieve 263 cm² V$^{-1}$ s$^{-1}$ at $V_{ds}$ = 1 V. Building on this initial result, we performed extensive device fabrication and repeated testing. Through these experiments, the highest recorded mobility reached 296 cm² V$^{-1}$ s$^{-1}$ with a ON-OFF current ratio above $10^6$, with the logarithmic form of transistor showed in insert figure of **Fig. 2a**. The morphology and the corresponding transfer curve of this device were provided in Supplementary **Fig. S3**, demonstrating the reliability and consistency of our fabrication process. Meanwhile, the output characteristics (**Fig. 2b**) displayed linear $I_{ds}$-$V_{ds}$ behavior over a range of $V_{ds}$ = 0~3 V, indicating the presence of ohmic contacts.

Compared to $WS_2$-FETs, $Ga_2O_3/WS_2$ transistor demonstrated significant improvements in both ON-state current and mobility. The comparison experiment, including transfer and output results, are displayed in Supplementary **Fig. S4**. All transistors showed the typical n-type characteristic. Notably, the ON-OFF ratio of $Ga_2O_3/WS_2$ transistor increased by an order of magnitude compared to the $WS_2$ transistor. The maximum $\mu_e$ increase from 68.67 cm² V$^{-1}$ s$^{-1}$ of $WS_2$ transistor to 212.84 cm² V$^{-1}$ s$^{-1}$ to $Ga_2O_3/WS_2$ transistor under ambient condition, and from 101.23 cm² V$^{-1}$

s$^{-1}$ to 253.97 cm$^2$ V$^{-1}$ s$^{-1}$ in vacuum. It is worth noting that the output curve of WS$_2$ transistor exhibited noticeable nonlinearity above $V_{ds}$ of 0.5 V, suggesting the formation of Schottky contact between multilayer WS$_2$ and deposited Au electrodes. In contrast, the Ga$_2$O$_3$ tunneling layer enabled a linear output response, confirming a significant reduction in the Schottky barrier and the establishment of ohmic contacts.

The improved electrical performance, including increased mobility and a highly linear output, is attributed to the tunneling contact provided by the Ga$_2$O$_3$ layer. Firstly, according to quantum mechanical principles, electrons can tunnel through an energy barrier even when their energy is lower than the barrier height. When a thin Ga$_2$O$_3$ layer is inserted between metal electrodes and WS$_2$, it attenuates the electron wave function of the metal, reducing its penetration into the semiconductor and thus lowering the SBH. Additionally, the interface between Ga$_2$O$_3$ and WS$_2$ forms a dipole due to differences in charge neutrality levels, preventing the built-in shift of Fermi levels and further reducing the Schottky barrier [17,21].

In addition to enhancing the electrical performance of transistors through tunneling contacts, the encapsulation function of Ga$_2$O$_3$ can be demonstrated by comparing device performance under ambient and vacuum conditions. Transconductance, a key performance index directly related to the amplification ability and current control efficiency of transistors, can be one representation to reflect this effect. **Fig. 2c** illustrates the variation in transconductance for WS$_2$ transistors and Ga$_2$O$_3$/WS$_2$ transistors under both ambient and vacuum conditions, with data extracted from Supplementary **Fig. S4c** and **S4f**. The Ga$_2$O$_3$/WS$_2$ transistor consistently achieved significantly higher transconductance across all on-state conditions, a result that aligned with the previously noted improvements in mobility. Moreover, transconductance measured in vacuum consistently exceeds that in ambient conditions for both transistors. Factors such as oxide etching, charge trapping, parasitic capacitance, and noise from air molecules usually affect transistors operating in ambient conditions [31,32]. This effect is particularly pronounced in transistors without encapsulation, explaining the substantial difference in transconductance for the WS$_2$ transistor, which increased from 3.1 mS in ambient to 13.5 mS in vacuum at $V_{gs}$ = 40 V, representing a 335.5% improvement. Therefore, an effective encapsulating layer is highly desired for the practical applications of 2D devices. For the $Ga_2O_3$/$WS_2$ transistor, where $Ga_2O_3$ also served as an encapsulating layer, protection from the influence of air molecules. The difference of transconductance could be mainly considered due to the variation of heat dissipation efficiency. Consequently, the increase in transconductance from ambient (29.1 mS) to vacuum (37.3 mS) was only 28.4%. These findings demonstrate the encapsulation effect of $Ga_2O_3$, ensuring stable transistor performance under ambient conditions.

**Fig. 2d** benchmarks the performance of the $Ga_2O_3$/$WS_2$ transistor, comparing its mobility and ON-OFF ratio against other reported values for multilayer $WS_2$ transistor. Our $Ga_2O_3$/$WS_2$ transistor demonstrated the highest mobility levels close to 300 $cm^2 V^{-1} s^{-1}$. This performance surpasses that achieved by other enhancement methods such as chemical doping [33–35] or plasma treatment [36] of $WS_2$ channel materials, dielectric layer design [37,38], and other contact engineering techniques [21,39–41]. Additionally, the ON-OFF ratio of our device is also relatively high compared to other $WS_2$-based transistors reported in the literature.

The transfer length method (TLM) measurement is typically used to measure the contact resistance of metal-semiconductor contacts. However, when conductors are separated by a thin insulating layer, the tunneling current and voltage along the contact length show spatial variations. The magnitude of tunneling resistance is directly related to the contact area in the longitudinal direction, where the larger contact areas result in lower tunneling resistance. In the transverse direction, the contact length becomes relevant after electrons tunnel through the tunneling layer. When the tunneling layer is uniform, sufficiently thin, and free of charge trapping effects, calculations can be simplified [42]. To facilitate performance comparisons with existing tunneling contact 2D transistors, we also employ the TLM method to characterize the specific contact resistivity of $Ga_2O_3$/$WS_2$ device, as the configuration shown in Supplementary **Fig. S5a**. The contact distances varied in a gradient from 2 μm to 28 μm. When current flows through the contact, each contact presents a voltage drop due to contact resistance, while current flow through the semiconductor creates a voltage drop due to body resistance. By plotting the resistance values against the contact distances, the contact resistance can be obtained from the Y-axis intercept, and the slope represents the body resistance. **Fig. 2e** shows the $I_{ds}$-$V_{ds}$ relationship for source-drain electrodes with different spacings in the $Ga_2O_3$/$WS_2$ device at $V_{gs}$ = 80 V. All measurements exhibited good linearity, allowing the total resistance value to be determined by fitting the slope of the *I-V* linear relationship. Then the specific contact resistivity for contact of $Ga_2O_3$/$WS_2$ transistor was extracted and the data are shown in **Fig. 2f**. Contact area of electrodes and $Ga_2O_3$ tunneling layer was defined as 20 μm x 10 μm. Finally, $\rho_c$ falls within the range of ~ 3-25 x $10^{-5}$ Ω·$cm^2$. This low resistivity could be attributed to the tunneling effect of the $Ga_2O_3$ layer and preventing Fermi level pinning with an excellent van der Waals contact. When compare to the traditional deposition metal electrode, the total contact resistance for the $WS_2$ transistor was significantly reduced from 32.52 kΩ·μm (Au electrodes) to 2.38 kΩ·μm ($Ga_2O_3$ tunneling contacts). The detailed fitting results for both deposited Au-contacts and $Ga_2O_3$ tunneling contacts are shown in Supplementary **Fig. S5b**. Consequently, the achieved contact resistance is within the range of ideal ohmic contacts and is among the lowest reported for $WS_2$ FETs with various contact engineering techniques (Supplementary **Fig. S5c**) [11,21,39–41,43]. It indicates the potential of $Ga_2O_3$ tunneling being applied to the actual use scenario of electronic devices.

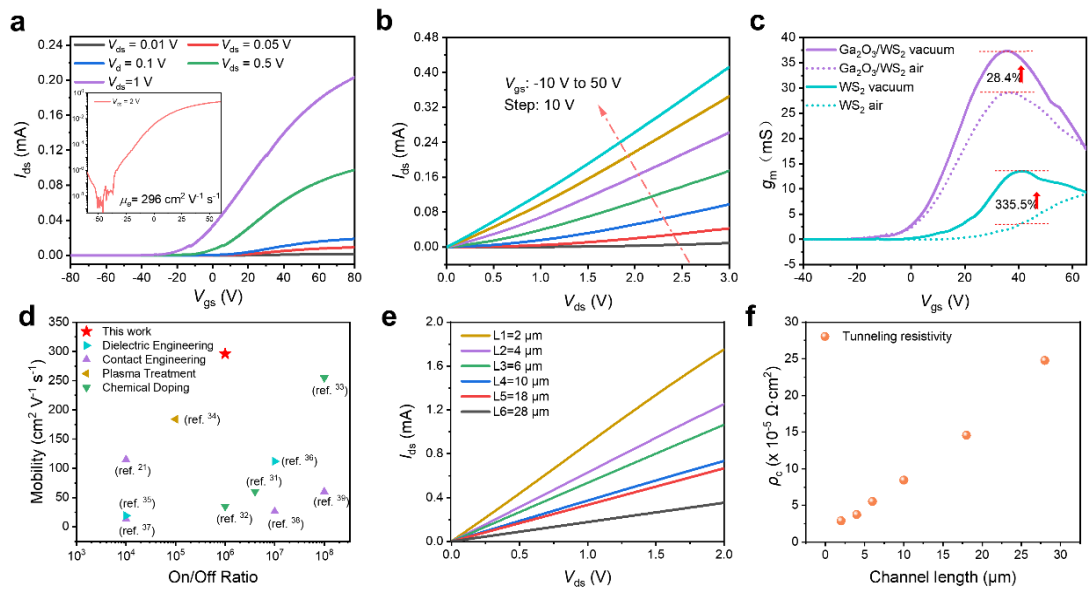

**Fig. 2| Room Temperature Electrical Characterization** (a) $I_{ds}$-$V_{gs}$ transfer

characterization of the $Ga_2O_3/WS_2$ tunneling contact transistor under vacuum conditions (~$10^{-4}$ Pa). Inset: The logarithm representation of transfer curve of the device with the recorded highest mobility, measured under $V_{ds}$ = 2 V. (**b**) $I_{ds}$-$V_{ds}$ output characterization of the $Ga_2O_3/WS_2$ tunneling contact transistor with $V_{gs}$ changes from -10 V to 50 V in step of 10 V. (**c**) Transconductance of $Ga_2O_3/WS_2$ tunneling device and $WS_2$ device under air condition and vacuum. $V_{ds}$=0.01 V. (**d**) Comparation of mobility with other multilayer $WS_2$ based FETs. (**e**) Resistance of the $Ga_2O_3/WS_2$ tunneling contact transistor channel at different contact distances. (**f**) Plot of the specific contact resistivity $\rho_c$ as a function of channel length.

**Barrier height comparison**

The contact resistance in transistors is primarily influenced by tunneling resistance and the barrier height. To gain deeper insights into the operating behavior of the $Ga_2O_3/WS_2$ transistor and to highlight its differences from pristine $WS_2$, we systematically conducted transport measurements at various temperatures. **Fig. 3a** and **3d** illustrate the temperature-dependent transfer characteristics of the $Ga_2O_3/WS_2$ and $WS_2$ transistors, respectively, as the temperature decreases from room temperature to 90 K. As the temperature decreases, the resistivity of the channel material decreases, primarily due to diminished phonon scattering, which enhances electron mobility [44]. In contrast, the contact resistance increases with cooling. According to the thermionic emission theory, lower temperatures make it more difficult for electrons to overcome the barrier from the metal contact into the semiconductor. This increase in contact resistance is particularly pronounced at low $V_{gs}$, where the lower concentration of charge carriers results in a higher dependence on the thermionic emission process for current flow [18]. At high $V_{gs}$, the increased gate bias induces a greater concentration of charge carriers at the semiconductor surface, which can help to mitigate the effect of the contact resistance by enhancing the tunneling effect and facilitating more efficient current flow. The interplay of these factors leads to complex temperature-dependent conductivity in transistors, often resulting in a metal-insulator transition (MIT) [45]. This principle explains the observed crossover from an insulating to a metallic regime, occurring around $V_{gs} \approx 60$ V for Ga$_2$O$_3$/WS$_2$ transistor and around $V_{gs} \approx 20$ V for WS$_2$ transistor.

The temperature-dependent mobility of both Ga$_2$O$_3$/WS$_2$ and WS$_2$ transistors, calculated from their metallic regimes, are presented in **Fig. 3b** and **3e**, respectively. As the temperature decreased from room temperature to 90 K, the field-effect mobility of the Ga$_2$O$_3$/WS$_2$ transistor showed a continuous increase, reaching approximately 1800 cm$^2$ V$^{-1}$ S$^{-1}$, while the WS$_2$ transistor only achieved 103 cm$^2$ V$^{-1}$ S$^{-1}$ at 90 K. This exceptionally high mobility at low temperatures is unprecedented among other methods [46,47]. In the high temperature regime, the mobility follows the expression $\mu_{FE} \propto T^{-\gamma}$, where lattice vibrations intensify, leading to more frequent electron-phonon scattering events. The temperature damping factor γ characterizes the severity of mobility dependency on temperature, which was found to be 1.32 for the WS$_2$ transistor and 2.48 for the Ga$_2$O$_3$/WS$_2$ transistor. This behavior was consistent across devices incorporating the same Ga$_2$O$_3$ tunneling layer, as demonstrated in Supplementary **Fig. S6**. The difference in the γ exponent between the Ga$_2$O$_3$/WS$_2$ and WS$_2$ transistors can be attributed to several factors. One possibility is the fundamental difference in charged-impurity scattering compared to WS$_2$ alone, leaving phonon scattering as the dominant mechanism at high temperatures [48,49]; Additionally, the encapsulation effect of the ultrathin Ga$_2$O$_3$ layer, with its superior interface engineering, minimizes structural damage and reduces scattering from material imperfections, contributing to the observed high mobility and unique temperature dependence.

To precisely determine the barrier height of the tunneling contact, Arrhenius plots (ln ($I_{DS}/T^{1.5}$) versus 1000/$T$) were extracted from the temperature-dependent transport measurements, as shown in Supplementary **Fig. S7a** and **S7b** for Ga$_2$O$_3$/WS$_2$ and WS$_2$ respectively. According to thermionic emission theory, the current in the subthreshold region is primarily governed by thermally assisted tunneling and thermionic emission, which can be described by the equation:

$$I_{ds} = A_{2D}^* T^{3/2} \exp\left(-\frac{q\phi_B}{k_B T}\right) \left[1 - \exp\left(-\frac{V_{ds}}{k_B T}\right)\right] \qquad (1)$$

where $A_{2D}^* = \frac{q(8\Pi k_B^3 m^3)^{0.5}}{h^2}$ is the Richardson constant for a 2D system, $h$ is the

Planck constant, $m$ is the electron effective mass, $T$ is the temperature, $k_B$ is Boltzmann's constant, $q$ is the elementary charge, and $\phi_B$ is the effective contact barrier height for a given gate-source voltage [50,51]. In Schottky barrier devices, the inverse subthreshold slope is influenced by thermally assisted tunneling, and the tunneling current becomes negligible only when the gate voltage drops below the flat band voltage ($V_{FB}$). This is crucial for accurately determining the barrier height at this specific gate voltage. By analyzing the slope of the curves in the high-temperature range using the simplified equation (2) derived from equation (1) [52], the gate voltage-dependent $\Phi_B$ values were collected and presented with error bars in **Fig.3c and 3f**.

$$\ln(I_{ds}/T^{1.5}) = \frac{-\Phi_B}{k_B T} + c \qquad (2)$$

At relatively low gate voltage, the effective barrier height displays a linear response to $V_{gs}$, indicating that the current is primarily dominated by thermionic emission. As the gate voltage increases, a deviation from this linear trend signifies the commencement of significant thermally assisted tunneling. The flat band voltage is determined by identifying the data point where the linear trend begins to deviate, making it possible to pinpoint the accurate barrier height determination. For the $Ga_2O_3/WS_2$ transistor, the barrier height at $V_{FB}$ was determined to be 3.7 meV, while for the $WS_2$ transistor, it was 258 meV. This substantial reduction in barrier height for the $Ga_2O_3/WS_2$ transistor, by two orders of magnitude validated the formation of ohmic contacts. This significant reduction could be attributed to the tunneling contact effect of $Ga_2O_3$ layer, thereby minimizing the interaction between the metal and the semiconductor and the forming of Schottky barrier. It not only attenuated the metal electron wave function but also mitigated the Fermi level pinning effect. Together, these mechanisms contributed to the superior electrical performance observed in devices incorporating the $Ga_2O_3$ tunneling layer.

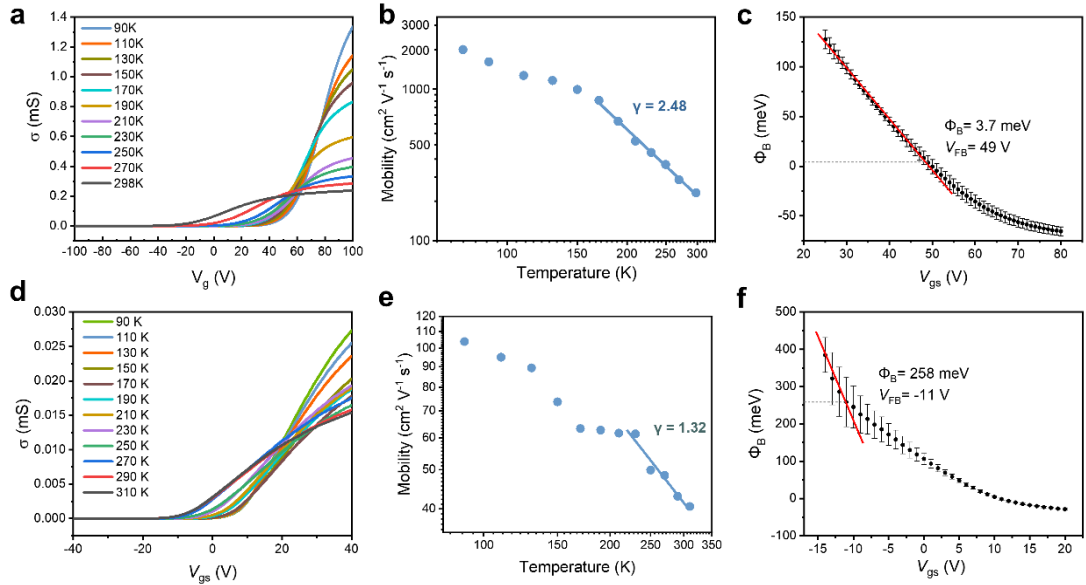

**Fig. 3| Temperature Dependence of Transport Characteristics.** Low temperature transfer curve of (**a**) $Ga_2O_3/WS_2$ tunneling contact transistor and (**d**) $WS_2$ transistor at $V_{ds} = 1$ V. Both of device exhibited metal-insulation transition. Temperature-dependent mobility of (**b**) $Ga_2O_3/WS_2$ tunneling contact transistor and (**e**) $WS_2$ transistor. At high temperature region is typically phonon scattering. The exponent γ was 2.48 for the tunneling contact and 1.32 for the Au-contact. Schottky barrier heights of the (**c**) $Ga_2O_3/WS_2$ tunneling contact transistor and (**f**) $WS_2$ transistor. The barrier height had largely reduction from 258 meV to 3.7 meV with the insert of $Ga_2O_3$ tunneling layer.

**Low temperature output and tunneling mechanism**

To gain a deeper understanding of the tunneling mechanism facilitated by the insertion of the ultrathin $Ga_2O_3$ layer, we examined the Fowler–Nordheim (F–N) tunneling and direct tunneling (DT) models in relation to the transport characteristics of the $Ga_2O_3/WS_2$ transistor. We first explored the key distinctions between of tunneling model, as the schematic energy band diagrams for both illustrated in **Fig. 4a**. According to quantum theory, if the potential barrier is sufficiently thin, the electron wave function can traverse it, leading to a non-zero probability of electrons appearing on the opposite side of the barrier. When a strong electric field is applied, F-N tunneling occurs, enabling the electron wave function to penetrate the triangular barrier and enter the conduction band of the dielectric material. Conversely, when the voltage across the insulating layer is low, electrons traverse the full thickness of the oxide, making the gate current dependent on direct tunneling [53]. It normally occurs with nano-thickness dielectric layer. In this context, the F-N and DT models can be described as follows:

F-N tunneling: $\ln\left(\frac{I_{ds}}{V_{ds}^2}\right) = \ln\left(\frac{Aq^3 m_0}{8\pi h \Phi_B d^2 m^*}\right) - \frac{8\pi\sqrt{2m^*}\Phi_B^{3/2} d}{3hqV_{ds}}$ (3)

Direct tunneling: $\ln\left(\frac{I_{ds}}{V_{ds}^2}\right) = \ln\left(\frac{Aq^2\sqrt{2m^*\Phi_B}}{V_{ds}h^2 d}\right) - \frac{4\pi\sqrt{2m^*\Phi_B}}{h}$ (4)

where $A$ is the electrical contact area, $d$ is the barrier width, $h$ is Planck's constant, $m^*$ is the effective mass of an electron in the WS$_2$ flake (0.33 $m_0$, where $m_0$ is the rest mass), and $\Phi_B$ is the barrier height [54].

The output curve of the Ga$_2$O$_3$/WS$_2$ transistor at $V_{gs}$ = 80 V continued to exhibit excellent linearity as the temperature decreases, as shown in **Fig. 4b**. This further verified the formation of an ohmic contact due to the Ga$_2$O$_3$ tunneling layer. Only a slight deviation from linearity was observed when the temperature drops below 100 K, which might be attributed to the increasing contact resistance at low temperatures. Based on the output results, a plot of $\ln(I_{ds}/V_{ds}^2)$ as a function of $1/V_{ds}$ was displayed in **Fig. 4c**. The curves exhibited a logarithmic dependence across all temperatures, rather than the linear dependence expected for F-N tunneling. This suggested that there was no significant evidence of F-N tunneling, even at higher voltage biases. Instead, DT Fitting demonstrated a linear relationship when $\ln(I_{ds}/V_{ds}^2)$ was plotted against $\ln(1/V_{ds})$, as exhibited in **Fig. 4d**, with the slope of the fitted line being close to 1. This nearly perfect fit to the DT model equation (4), indicating that the charge carrier injection in the Ga$_2$O$_3$/WS$_2$ transistor was dominated by direct tunneling. Direct tunneling leverages quantum effects and can play a crucial role in emerging technologies such as quantum computing and quantum communication. Unlike F-N tunneling, direct tunneling does not rely on high electric fields to reduce the barrier, which allows devices to operate with lower power consumption, faster response times, and reduced heat generation. This explains the exceptional electrical performance of the Ga$_2$O$_3$/WS$_2$ transistor with the Ga$_2$O$_3$ tunneling layer and highlights its potential for future applications in advanced electronic circuits. By applying the fitting equation, the DT barrier parameter ($d\sqrt{\Phi_B}$), which quantifies the ease with which electrons can tunnel through the $Ga_2O_3$ barrier, was found to vary in the range of 0.025 - 0.032 $meV^{1/2} \cdot nm$ across the temperature range of 77.5 to 300 K, as shown in Supplementary **Fig. S8**. Although the $Ga_2O_3$ tunneling layer was thicker than previously studied hBN layers, the barrier parameter for $Ga_2O_3$ tunneling contacts is significantly lower than that of hBN [21]. This relatively low range indicated that the $Ga_2O_3$ tunneling layer provided a more efficient tunneling pathway compared to other materials like hBN or metal formed junctions [55], thereby contributing to the observed high performance of the $Ga_2O_3/WS_2$ transistor. It further confirmed that the addition of the $Ga_2O_3$ tunneling layer leads to a substantial reduction in the barrier height.

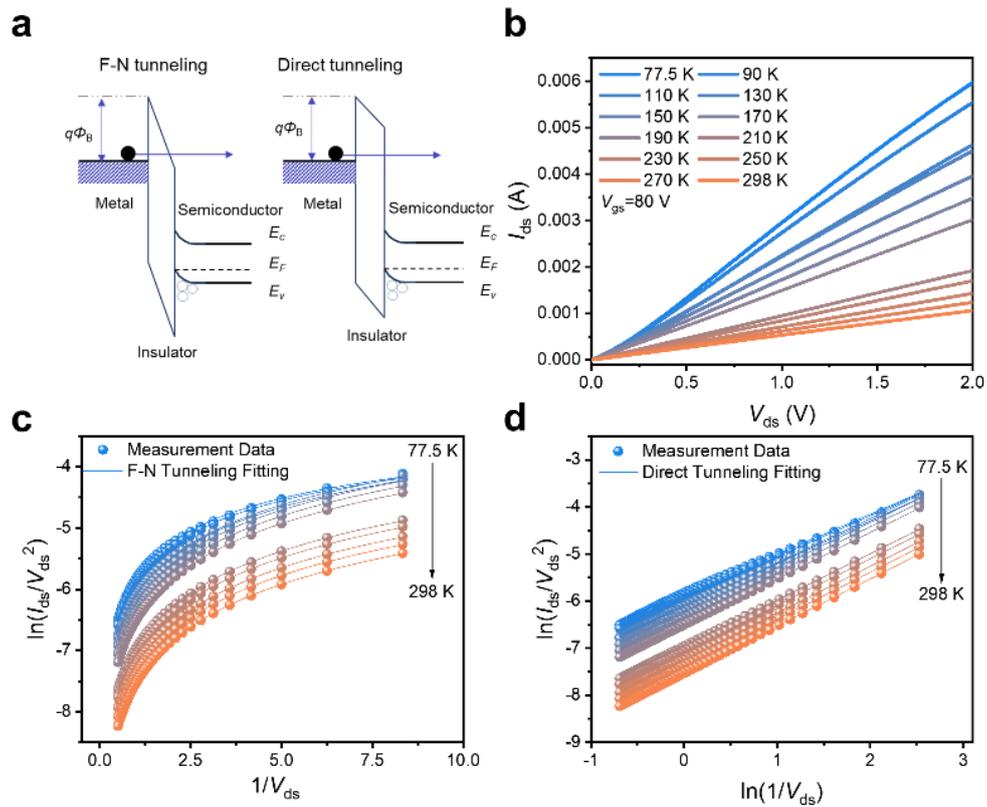

**Fig. 4| Tunneling Mechanism Analysis.** (**a**) The schematic energy band diagrams for Fowler-Nordheim tunneling and direct tunneling. (**b**) Output curve with temperature decrease. Linear output displayed at all temperature especially above 90 K. (**c**) Fowler-Nordheim tunneling fitting with temperature dependent output data. Dash lines show logarithmic relations fitted to the data. (**d**) Direct tunneling fitting with temperature dependent output data. Dash lines display good linear fit.

**Scalability and Reproducibility of Ga$_2$O$_3$/WS$_2$ FETs**

Building upon the low-cost, large-area scalable fabrication of ultrathin Ga$_2$O$_3$ layer, we explored the potential for creating scalable transistor arrays on single chip. By leveraging the uniform and high-quality properties of the printed Ga$_2$O$_3$ films, we successfully fabricated an extensive array of Ga$_2$O$_3$/WS$_2$ FETs, demonstrating the feasibility of this approach for large-scale device integration [30]. The following sections detail the experimental results obtained from these arrays which exhibit impressive performance metrics and consistency across a wide substrate area. **Fig. 5a** illustrated the layout of Ga$_2$O$_3$/WS$_2$ FET arrays fabricated on a 1.5 x 1.5 cm² SiO$_2$/Si chip. The multilayer WS$_2$ nanoflakes, obtained through exfoliation, had controlled thickness less than ten layers to ensure uniform electrical properties of the channel material. These nanoflakes were evenly distributed across the entire substrate. A large-area Ga$_2$O$_3$ film was then covered on all WS$_2$ nanoflakes, followed by deposition of Au electrodes.

The electrical characteristics of the Ga$_2$O$_3$/WS$_2$ FETs were displayed in **Fig. 5b** and **5c** with the transfer curve measured under a $V_{ds}$ of 1 V. The linear transfer characteristics, where the $I_{ds}$ increased linearly with the $V_{gs}$ over a wide range, indicated the formation of high-quality ohmic contacts and effective gate modulation for all transistor arrays. Meanwhile, the logarithmic representation of the transfer characteristics revealed the good repeatability of high ON-OFF current ratio (all exceeding $10^6$), which is crucial for digital switching applications. The experimental results indicate that channel material was heavily doped, primarily due to the intrinsic quality of the WS$_2$, which also leads to the higher subthreshold slope. This observation is consistent with the transfer characteristics measured separately for WS$_2$ (Supplementary **Fig. S4**). Additionally, most encapsulated WS$_2$ devices maintained stable performance for over three weeks, with less than 20% reduction in mobility, as shown in **Fig. 5d**. The variation of mobility $\Delta\mu$ was defined as $\mu-\mu_0$, where $\mu_0$ was the initial mobility and $\mu$ was for the mobility measured after a period of time. The error bar was calculated with the statistics of 11 devices, ensuring the reliability of our stablity measurement. The variation of ON-OFF ratio also maintain within an order of magnitude after three weeks, indicating that good switching characteristics were still retained in Ga$_2$O$_3$/WS$_2$ transistor. These results demonstrated the effectiveness of $Ga_2O_3$ encapsulation in preserving the electrical properties of $WS_2$ over time while acting as a tunneling layer.

The electrical parameter of 25 transistors, including mobility and ON-OFF ratio, were collected and statistically analyzed. **Fig. 5e** presented a histogram of mobility, which followed a relatively Gaussian distribution, with an average $\mu_e$ of 191 cm² V⁻¹ s⁻¹ ± 43.5 cm² V⁻¹ s⁻¹. The standard deviation (σ) and correlation coefficient ($R^2$) from the fitting process supported the high confidence of statistics. All achieved mobilities were significantly higher than those typically reported for $WS_2$ transistors without tunneling contacts. It demonstrated the reproducibility and reliability of the $Ga_2O_3$ tunneling contact approach, which is essential for scalable electronics. The corresponding ON-OFF ratio distribution, with most devices exhibited values between $10^6$ and $10^8$ (**Fig. 5f**). Such high ON-OFF ratios are indicative of excellent switching performance, being appropriate for low-power and high-speed electronic applications. These statistical results validated the scalability and reproducibility of $Ga_2O_3$ as a tunneling contact layer, demonstrating its practical value and significance for industrial and commercial applications in 2D semiconductor electronic devices. Specifically, the printed $Ga_2O_3$ layer offers significant advantages over other tunneling layer materials in integrating with large-area grown $WS_2$ nanolayer. This synergy enables the formation of industrial-scale device arrays, where $Ga_2O_3$ serves as an efficient tunneling contact layer. Its compatibility with large-scale $WS_2$ film deposition techniques facilitate the production of uniform and high-performance electronic devices, making it a promising candidate for the development of scalable and commercially viable semiconductor technologies.

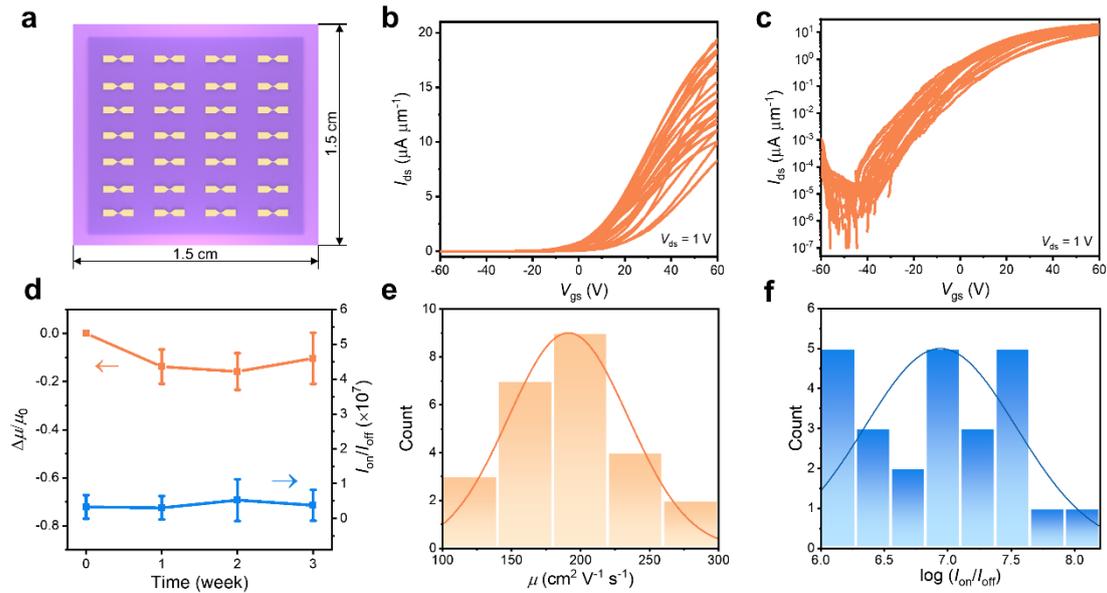

**Fig. 5| Statistics of scaled devices.** (a) Schematic diagram of $Ga_2O_3$/$WS_2$ FET arrays fabricated on 1.5 x 1.5 $cm^2$ $SiO_2$/Si chip. Deep purple region represents for $Ga_2O_3$ layer. (b) linear representation and (c) Logarithmic representation of the transfer characteristics of $Ga_2O_3$/$WS_2$ FETs with $L_{CH}$ of 4.5 μm and $L_{CW}$ of 6 μm, under $V_{ds}$ = 1 V. (d) Long time stability of FETs for three weeks measurement. The variation of mobility $\Delta\mu=\mu-\mu_0$. The variation of ON-OFF ratio after weeks were presented as $I_{on}/I_{off}$. The error bar represent for statistics of devices. (e)Histograms showing the variation in electron mobility for $Ga_2O_3$/$WS_2$ FETs. A total of 25 transistors were tested and counted. (f) Histograms of corresponding ON/OFF ratio of transistors.

## Conclusion

In this study, we have demonstrated the success of ultrathin amorphous $Ga_2O_3$ layer as tunneling contact layers for $WS_2$-based FETs. Our findings revealed that the incorporation of $Ga_2O_3$ significantly reduced the barrier height from 258 meV to 3.7 meV, approaching ideal ohmic contact conditions. The $Ga_2O_3$ layer also acted as an effective passivation layer, protecting the $WS_2$ channel from environmental contaminants and maintaining device stability. The $Ga_2O_3$/$WS_2$ transistors exhibited superior electrical performance, including ultra-high electron mobility of up to 296 $cm^2$ $V^{-1}$ $s^{-1}$ at room temperature and low specific contact resistivity within range of ~ 3-25 x $10^{-5}$ Ω·cm$^2$. These results surpassed those obtained from traditional contact engineering methods. Different from other tunneling layer, the scalable and low-cost fabrication of Ga$_2$O$_3$ films is particularly advantageous for practical electronic device manufacturing. Overall, the introduction of Ga$_2$O$_3$ as a tunneling contact layer not only addresses the challenges of high contact resistance and poor environmental stability in 2D semiconductor devices at the same time, but also opens new avenues for the development of high-performance, scalable, and industrial-grade electronics.

## Methods

**Materials Preparation**

WS$_2$ nanoflakes were mechanically exfoliated from a single crystal (purchased from HQ Graphene) and transferred onto a SiO$_2$/Si substrate with a 300 nm thick SiO$_2$ layer. A printer head was assembled using two glass slides with a 2 mm gap. Liquid gallium (Sigma-Aldrich, 99.99% trace metal basis) sources were melted and injected into the printer head using a syringe. PPC substrates were prepared by spin coating the solution onto the desired PDMS substrates. The printer head was positioned close to the target substrate until the liquid metal meniscus contacted the surface. The printing process was initiated by translating the substrate at a controlled speed to achieve uniform printing. The preparation process of large area Ga$_2$O$_3$ was shown in the appendix video.

**Device Fabrication**

A clean, continuous region of the Ga$_2$O$_3$ layer, free from gallium residues, was selected. Using a dry-transfer platform, the Ga$_2$O$_3$ film was transferred onto the WS$_2$ nanoflakes with the aid of PPC layer as the transfer medium. The residual PPC was removed using acetone, leaving WS$_2$ nanoflakes covered with the Ga$_2$O$_3$ layer. Ga$_2$O$_3$/WS$_2$ back-gated FETs were fabricated using photolithography to pattern the electrodes, followed by the deposition of 50 nm thick Au electrodes through thermal evaporation and lift-off processes.

**Materials Characterization**

The thickness of each layer was measured using AFM (Bruker/Dimension Icon). Optical properties were analyzed using Raman spectroscopy and photoluminescence spectroscopy (LabRAM HR800, Horiba Jobin-Yvon) with a 532 nm excitation wavelength. Laser power was adjusted to approximately 1 mW to ensure consistent excitation conditions for samples exposed to air for different durations. The signals were captured for 1 s and accumulated over three measurements. XPS analysis was performed on $SiO_2$/Si substrates using a Thermo Scientific K-Alpha XPS spectrometer, equipped with a monochromatic Al Kα source and a concentric hemispherical analyzer (CHA) to determine the chemical composition.

**Electrical Characterization**

All electrical characterizations were performed at room temperature in a high-vacuum environment (~$10^{-4}$ Torr) using a Keysight B1500A semiconductor parameter analyzer. Standard DC sweeps were used to measure the transfer and output characteristics of all devices. To stabilize the FETs, each device underwent multiple repetitions of the same measurement. Transfer characteristics were measured three times for conditioning, with the fourth measurement used for analysis. Output characteristics were measured twice after the transfer characteristics, with the second measurement used for analysis.

**Acknowledgements**

Y. Li. and T. Yun contributed equally to this work. We acknowledge the support from the National Natural Science Foundation of China (No. 62204165). Guangdong Basic and Applied Basic Research Foundation (No. 2021B1515120034), National Key R&D Program of China (No. 2021YFA1202902), the National Natural Science Foundation of China (No. 12204336), and Songshan Lake Materials Laboratory (No. Y0D1051F211).

**Conflict of Interest**

The authors declare no conflict of interest.

**Data Availability Statement**

The data that support the findings of this study are available in the supplementary material of this article.

**Supplementary Information**

The supporting information is available free of charge.

Figures S1 to S8: representative demonstration for the tunneling layer transfer process, electrical measurement, tunneling barrier calculation and so on. (PDF)

Supplementary video: Printing process of $Ga_2O_3$ tunneling layer.